# Synthesizing Diabetic Foot Ulcer Images with Diffusion Model

Reza Basiri[1,2][0000-0002-0209-6478], Karim Manji[3,4], Harton Francois[3,4], Alisha Poonja[3,4], Milos R. Popovic[1,2][0000-0002-2837-2346], Shehroz S. Khan[1,2][0000-0002-1195-4999]

[1] Institute of Biomedical Engineering, University of Toronto, Canada
[2] KITE, University Health Network, Canada
[3] Zivot Limb Preservation Centre, Peter Lougheed Centre, Canada
[4] Department of Surgery, Cumming School of Medicine, University of Calgary, Canada

Corresponding Author: Reza Basiri
Reza.basiri@mail.utoronto.ca

**Abstract.** Diabetic Foot Ulcer (DFU) is a serious skin wound requiring specialized care. However, real DFU datasets are limited, hindering clinical training and research activities. In recent years, generative adversarial networks and diffusion models have emerged as powerful tools for generating synthetic images with remarkable realism and diversity in many applications. This paper explores the potential of diffusion models for synthesizing DFU images and evaluates their authenticity through expert clinician assessments*. Additionally, evaluation metrics such as Fréchet Inception Distance (FID) and Kernel Inception Distance (KID) are examined to assess the quality of the synthetic DFU images. A dataset of 2,000 DFU images is used for training the diffusion model, and the synthetic images are generated by applying diffusion processes. The results indicate that the diffusion model successfully synthesizes visually indistinguishable DFU images. 70% of the time, clinicians marked synthetic DFU images as real DFUs. However, clinicians demonstrate higher unanimous confidence in rating real images than synthetic ones. The study also reveals that FID and KID metrics do not significantly align with clinicians' assessments, suggesting alternative evaluation approaches are needed. The findings highlight the potential of diffusion models for generating synthetic DFU images and their impact on medical training programs and research in wound detection and classification.

*WoundVista: http://bit.ly/WoundVista

# 1 Introduction

Diabetic Foot Ulcer (DFU), a type of skin wound commonly found on the plantar foot, requires specialized care. Failure to heal DFUs can lead to prolonged treatments with high recurrence rates, resulting in the loss of mobility, independence, quality of life, amputation, and even mortality [1–3]. The global prevalence of DFUs is reported to be 6.3%, while in North America, it is estimated to affect 13% of the diabetic patient population [4]. In evaluating DFUs, clinicians follow a series of steps that involve recording and examining the location and size of the wounds, either in person or by analyzing captured images [5]. The interpretation of DFU images is a crucial skill for expert clinicians, enabling them to track and diagnose wound conditions and conduct research. However, the availability of real DFU datasets is limited, restricting clinical training and hampering research activities in the DFU area [6].

In recent years, there has been a notable emergence of Generative Adversarial Networks (GANs) [7] and diffusion models [8], which have proven to be powerful tools for generating synthetic images that exhibit remarkable realism and diversity in many image-processing applications [9, 10]. These techniques hold great promise, especially in medical images, as they can significantly enhance medical training programs and expedite research efforts.

The traditional approach to collecting medical images involves a laborious and time-consuming gathering of real-life samples. Due to medical record privacy and ownership concerns, this often entails redundant data collection from multiple centers. This conventional approach is burdened with limitations, including limited dataset size, imbalanced representation, data sharing constraints, and high costs associated with data acquisition. However, GANs and diffusion methods offer an alternative that overcomes these limitations, facilitating medical analysis and management advancements.

Although synthetic medical images have been extensively explored in areas like X-rays [11] and MRIs [12], the potential for generating synthetic wound images still needs to be explored, particularly in DFU. The availability of a large and diverse DFU dataset holds significant value for medical training programs and the development of standardized wound detection and classification methods. By encompassing various types of wounds, different severities, and distinct anatomical locations, an extensive collection of synthetic wound images enhances the representativeness of the dataset. This, in turn, facilitates the training of deep learning algorithms, allowing them to generalize better and recognize the diverse patterns of wounds encountered in clinical practice. While GNAs pioneered the area of synthetic image generation, diffusion models have risen in popularity due to their ability to generate higher-resolution images, a capability that is crucial in medical imaging.

In this study, we explore the potential of diffusion models for generating synthetic DFU images. To evaluate the authenticity of the synthetic DFUs, three expert clinicians assess and distinguish them as either real or fake. Additionally, we investigate the suitability of commonly used evaluation metrics such as Fréchet Inception Distance (FID) and Kernel Inception Distance (KID) for evaluating the quality of synthetic DFU image generation.

## 2 Related Works

Diffusion models involve introducing noise to data and then removing the noise iteratively to recover the original data. These models excel at capturing intricate relationships between signals and noise, leading to improved accuracy in reconstructing images. Moreover, diffusion models fall under probabilistic distributions. They can be conditioned to generate synthetic data of exceptional diversity and quality, making them particularly well-suited for image generation tasks. Although diffusion models hold promise for various applications, customizing them for specific domains, such as DFU image generation, may necessitate additional modifications. Having said this, diffusion models are relatively uncommon in medical imaging, as GANs remain the dominant architecture for generating images from noise or transforming images [13].

Rombach et al. [10] introduced a novel approach called latent diffusion models, where the forward and reverse processes occur in the latent space learned by an autoencoder. They enhanced the architecture by incorporating cross-attention, improving conditional image synthesis, super-resolution, and inpainting. Building upon Rombach's work, Packhäuser et al. [11] utilized the latent diffusion model to generate high-quality chest X-ray images conditioned on specific classes. They also proposed a sampling strategy to preserve the privacy of sensitive biometric information during the image generation process. The generated dataset was evaluated in a thoracic abnormality classification task, and the results demonstrated superior performance compared to GAN-based methods. Our approach to using an unconditional diffusion model takes inspiration from Rombach et al. [10] and Packhäuser et al. [11]'s works.

## 3 Method

### 3.1 Dataset

We used 2,000 640x480 DFU colored images from Yap et al. [14, 15] for training the diffusion model. Wound quantities per image ranged from zero to five in this dataset. The proportion of the wound area ranged from 0.06% to 57.38% of the entire image area. Some real DFU images are shown in Figure 1.

### 3.2 Synthetic Generator

GAN models have limitations when it comes to generating high-resolution images. To generate synthetic DFU imaging that matches the resolution of our DFU dataset, we chose to utilize an unconditional diffusion model. Our diffusion model was inspired by the work of Rombach et al. [10], with modifications made to accommodate our specific data size and computational resources.

During the training process, the diffusion model was trained on 256x256 input images of the entire foot, ensuring that at least one DFU was present in the image, along with minor background details. A batch size of 32 and an initial decaying learning rate of 1e-4 were used for approximately 500 epochs. Gaussian noise was added to the original image in the forward pass of diffusion. A denoising diffusion probabilistic model scheduler was employed to define the variance schedule across 1,000 successive steps. For the denoising step, a U-Net architecture with attention layers was utilized to predict the noise distribution and project it back to the original DFU images. Mean squared error loss was employed for this process.

In the inference stage, 1,000 denoising steps were applied in a reverse pass to generate 50 synthetic DFU images. These generated images comprised the foot, minor background elements, and DFUs on the foot. The 256 by 256 generated images were manually cropped to 100 by 100 boxes, ensuring the DFUs were centered in each box.

### 3.3 Analysis

We replicated the process of cropping 100 by 100 boxes from real DFU images to match the dimensions of the synthetic images. We employed FID, KID, and the ratings of three expert DFU clinicians to evaluate the similarity between the synthetic and real normalized images. FID and KID are metrics that compute the pixel-based similarity between the distributions of synthetic and real images. FID was calculated using 2048 Inceptionv3 feature layers, while KID utilized a subset sampling of 50 images.

For the clinicians' evaluation, we presented 100 cropped DFU images at 72 DPI, evenly split between real and synthetic, to three DFU clinicians (co-authors #2,3,4). Each clinician marked each image as either real or fake. We then compared and assessed the clinicians' markings and the correlation between FID, KID, and the clinicians' ratings using the student t-test and Pearson correlation coefficient, respectively.

All values, including FID, KID, ratings of real images, and ratings of synthetic images, were normalized to a distribution ranging from 0 to 3. A rating of 0 indicated no correct marking. In contrast, a rating of 3 indicated that all three clinicians correctly identified whether the image was a real or synthetic DFU, as shown in Fig 1.

## 4 Results

### 4.1 Generated Images

Synthetic images were generated from randomized Gaussian noises. The results were curated, and those with obvious color offsets of a real wound or foot were discarded. Fig 1 contains some sample synthetics images that were discarded and selected as well as some real DFU images for comparison. For illustration, we also show images before the 100 by 100 cropping application in Figure 1.

### 4.2 Evaluations

Of the 100 presented images, 77% were marked as real by the clinicians. Out of 300 ratings for the three clinicians, 84% of times marked real images correctly. This value falls to 30% for synthetic image correct marking. This means that 70% of the times synthetic DFU images were marked as real. The two marking distributions are significantly different, with a p-value of 0.01. On average, as shown in Fig1, 2.52 ±0.70 clinicians marked real DFU images as real, and 2.10 ±0.88 clinicians marked synthetic DFU images as real.

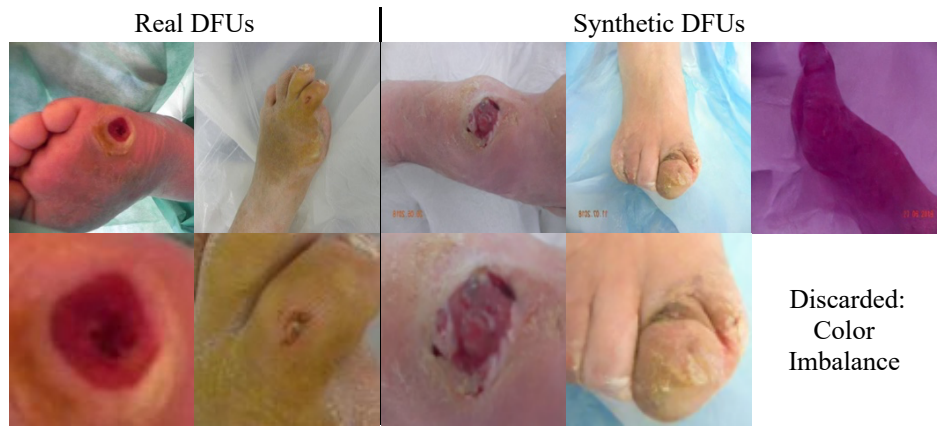

**Fig 1. Sample real and synthetic DFU images.** The original real and generated DFU images are depicted on the top row and 100 by 100 cropped 72 DPI versions are on the bottom. Generated images with color imbalances as shown in the last column were discarded.

Our mean values for the normalized FID and KID were 0.73 and 0.14. Pearson's correlation between FID and KID and clinicians' markings showed a positive coefficient of 0.20 but was not significant. The clinicians' ratings are shown in Fig 2.

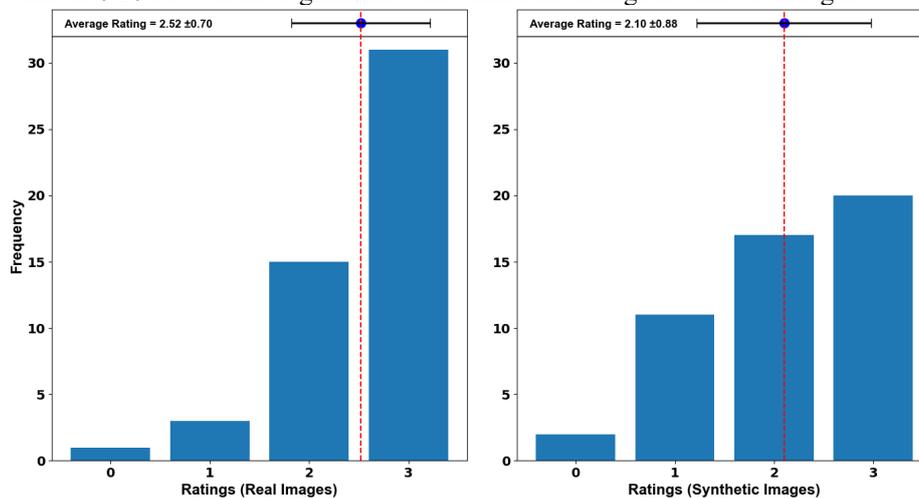

**Fig 2. Three clinicians' ratings of real and synthetic DFU images.** A rating of 0 means zero real rating for real or synthetic images from the three clinicians. While a noticeable number of synthetic images were marked as real, the confidence level in real DFU markings was higher, meaning clinicians were more frequently in agreement with the real DFUs.

## 5  Discussion

### 5.1  Application in the DFU Area

In medical imaging, high-resolution images that accurately depict local information is crucial for effective disease detection. Diffusion models have demonstrated superior performance compared to GANs in achieving this objective, leading to their increasing popularity. In our study focusing on DFUs, the diffusion model showed satisfactory performance despite having a limited dataset. With minor adjustments to the generated images, our current network successfully augmented the existing DFU datasets, enhancing their size and diversity. As evident from clinicians' ratings, the diffusion model synthesized wound images visually indistinguishable from real wounds. However, the confidence level in identifying real images was significantly higher, resulting in noticeably different rating distributions with a p-value of 0.01 for the student t-test. This disparity can be attributed to the small training dataset and the limited resolution of our model. This issue is resolvable by more training samples and expanding the computational power to train larger diffusion models for higher-resolution outputs.

Interestingly, we observed that commonly used evaluation metrics such as FID and KID did not significantly align with the assessments provided by clinicians. We discovered a statistically insignificant positive correlation of 0.20 for the Pearson r coefficient between these metrics and the ratings assigned by the clinicians. This suggests that alternative quantitative approaches for evaluating synthetic wound images may be required to better align with the clinicians' assessments.

### 5.2  Limitations and Future Directions

The medical community places significant importance on safeguarding the privacy of medical data. Some research findings indicate that diffusion models tend to memorize individual images from the training data and reproduce them during the generation process [16]. Consequently, adversaries could exploit this behavior to extract sensitive training data. Moreover, these studies reveal that diffusion models exhibit lower privacy than generative models like GANs. Therefore, there is a need for new advancements in privacy-preserving training methods to address these vulnerabilities, especially in sensitive health domains, such as medicine.

Utilizing diffusion models as a generative prior can help mitigate data heterogeneity, reduce the risk of privacy breaches, and enhance the quality and trustworthiness of the learned models in terms of fairness and generalization. Additionally, due to similar privacy concerns, diffusion models can streamline the process of generating synthetic medical data for educational purposes or augmentation while maintaining privacy and data integrity.

Even at the current stage, the diffusion model can generate images to enhance the existing dataset augmentations. Overfitting is a well-known problem with classification models, especially in DFU, where datasets are limited [6]. The standard solution to reduce overfitting is data augmentation that artificially enlarges the dataset. Classic augmentation techniques mostly include affine transformations such as translation, rotation, scaling, flipping and shearing [17]. Synthetic DFUs can complement classic augmentation by adding newly diffused DFU examples created by the diffusion model. As a next step, we will explore the effect of diffused DFU augmentation on classification model performances. Our team also made a web application called "WoundVista", publicly available for researchers to rate DFU images as real or fake at [http://bit.ly/WoundVista]; a screenshot is shown in Fig 3. We will use global and clinical performance in rating synthetic and real images to compare and analyze generative quantitative metrics such as FID or KID in DFU image evaluations.

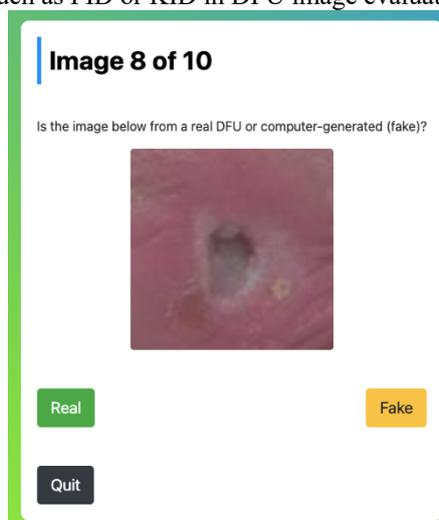

**Fig 3. Wound Rater Application.** We are collecting evaluations on our generated DFU images from clinicians and the public by using our web application. Global and clinical performance of real and synthetic DFU images are used to evaluate quantitative generative evaluation metrics such as FID.

## 6   Conclusion

We investigated the applications of diffusion models for generating synthetic DFU images. By generating diverse and realistic synthetic DFU images, these models can address the limitations of traditional data collection methods, such as limited dataset size, imbalanced representation, and high costs associated with data acquisition.

Our findings demonstrated that the diffusion model successfully synthesized DFU images that were visually indistinguishable from real wounds. Additionally, we demonstrated that FID and KID evaluation metrics did not align with the clinicians' assessments. This suggests the requirement for alternative quantitative approaches that better align with the expert clinicians' evaluations.

The utilization of diffusion models in generating synthetic DFU images shows promise for augmenting existing datasets and facilitating advancements in DFU analysis.